\documentclass[times, twoside]{zHenriquesLab-StyleBioRxiv}
\usepackage{amsmath} 
\usepackage{graphicx} 

\leadauthor{Zhuoyan Shen} 

\begin{document}

\title{Pan-Cancer mitotic figures detection and domain generalization: MIDOG 2025 Challenge}
\shorttitle{UCL Submission for MIDOG 2025}


\author[1]{Zhuoyan Shen}
\author[1]{Esther B\"ar}
\author[1, 2]{Maria Hawkins}
\author[3]{Konstantin Br\"autigam\textnormal{\textsuperscript{\textdagger}}\thanks{\textnormal{\textsuperscript{\textdagger}}These authors contributed equally.}}
\author[1]{Charles-Antoine Collins-Fekete\textnormal{\textsuperscript{\textdagger}}}

\affil[1]{Department of Medical Physics and Biomedical Engineering, UCL, UK}
\affil[3]{Centre for Evolution and Cancer, The Institute of Cancer Research, London, UK}
\affil[2]{Department of Radiotherapy, University College London Hospitals, London, UK}

\maketitle

\begin{abstract}
This report details our submission to the Mitotic Domain Generalization (MIDOG) 2025 challenge, which addresses the critical task of mitotic figure detection in histopathology for cancer prognostication.  Following the "Bitter Lesson"\cite{sutton2019bitterlesson} principle that emphasizes data scale over algorithmic novelty, we have publicly released two new datasets to bolster training data for both conventional \cite{Shen2024framework} and atypical mitoses \cite{shen_2025_16780587}. Besides, we implement up-to-date training methodologies for both track and reach a Track-1 F1-Score of 0.8407 on the MIDOG++ test set, as well as a Track-2 balanced accuracy of 0.9107 for atypical mitotic cell classification.
 
\end{abstract}


\begin{corrauthor}
c.fekete@ucl.ac.uk
\end{corrauthor}
\section*{Introduction}
Quantification of mitotic activity is a cornerstone of grading numerous types of cancers, including breast cancer, sarcomas, neuro-endocrine tumors, melanoma, and many other tumors \cite{elston1991}. However, manual identification of mitotic figures is a laborious and subjective task, suffering from high variability between observers \cite{vandiest2004}. Automated methods that use deep learning offer a promising solution, but often struggle with domain shift: variations in staining, scanning, and tissue preparation in different laboratories. The MIDOG challenge series \cite{Aubreville2023, AUBREVILLE2023102699} specifically targets this problem by evaluating algorithms on unseen data sources. In this work, we present our solution to both tracks of the latest MIDOG 2025 Challenge.

\section*{Track 1: Robust Mitosis Detection}

\subsection*{Material and Methods}
The OMG-Octo database \cite{shen_2024_14246170} was used for developing the models. Differently from our recent work~\cite{Shen2024framework}, the MIDOG 2022 test set was merged into our training set for optimal accuracy. Our method employs an ensemble of YOLOv10 object detection models applied to patches generated via a sliding window approach. Yolov10 features Consistent Dual Assignments for one-to-many (during training) and one-to-one (during inference) label assignment which preserves performance without broad non-maximum suppression, enhancing efficiency, which allowed us to run multiple models within the time allocated. 

During training, we augmented the data via colour jittering (hue ±1.5\%, saturation ±70\%, brightness ±40\%), random translation (10\% of image size), scaling (±50\%), horizontal flipping (p = 0.5), mosaic augmentation (combining four images), RandAugment, and random erasing (p = 0.4). In addition, we implemented histology-specific colour augmentation for H\&E-stained images, where RGB images were deconvolved into H\&E channels using stain vectors \cite{coloraug1}, followed by perturbation of stain concentrations via uniform sampling before reconstruction into RGB space~\cite{coloraug2}.

To enhance model robustness and improve recall, we utilize a Test-Time Augmentation (TTA) strategy where, for each patch, the model performs predictions on horizontal and vertical flip of the images, as well as the original. All prediction are averaged onto a single prediction test sets.


Beside TTA, we have trained 5 model on 5 different folds of the train/validation dataset.  For prediction, each submodel is ran through the TTA-augmented test images. The TTA-predictions are averaged to provide a unique set of predictions per model. To consolidate these into a final set of unique predictions, we use the Weighted Boxes Fusion (WBF) algorithm \cite{wbf}. Unlike traditional Non-Maximum Suppresion (NMS), which discards all but the highest-scoring box in an area, WBF combines the boxes in a cluster by averaging their coordinates and scores, weighted by their confidence. To account for the number of ensemble models, the final confidence score $C$ is re-scaled based on the number of boxes in the cluster ($T$) and the total number of models ($N$):
$$C_{final} = C_{avg} \times \frac{T}{N}$$
This formula acts as a consensus mechanism. The best confidence threshold is found by computing the F1-score on the validation sets at various thresholds.

\begin{figure}[!h]
    \centering
    \includegraphics[width=0.5\linewidth]{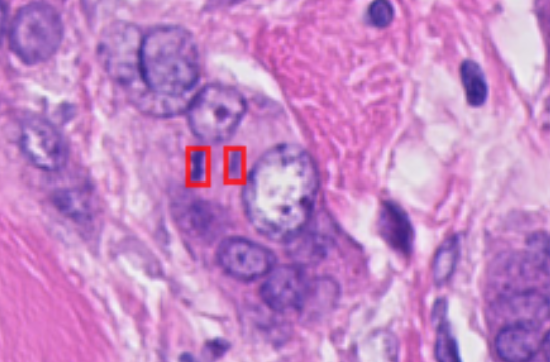}
    \caption{Mitotic cell in telophase, which is predicted as two individual mitotic cell by the object detection algorithm.}
    \label{fig:telophase}
\end{figure}

Finally, a key challenge identified through visual inspection is the dual detection of a single telophase mitotic cell as shown on Fig.~\ref{fig:telophase}.

Pathologists annotate this process as a single cell division event, but our model often predicts two separate bounding boxes for the daughter nuclei. Since these non-overlapping boxes fall outside the scope of standard NMS, we implemented a secondary, distance-based aggregation step. We merge bounding boxes whose centroids are within 10 $\mu$m, a threshold chosen both empirically, as it maximises the F1-score, and biologically, as it corresponds to the average cell radius (5-10 $\mu$m), consolidating the separate daughter nuclei predictions into a single mitotic cell

\subsection*{Preliminary Results}
Preliminary results on the validation set for each models demonstrate an ensemble F1-score equivalent to that of our OMG-Net report, \textit{i.e.} 0.8407 on canines and human specimens, and of 0.8391 on human specimen only. 

\section*{Track 2: Atypical Mitosis Classification}
\subsection*{Material and Methods}
We used four open-source datasets, AMi-Br, the MIDOG 2025 Atypical Training Set, LUNG-MITO \cite{ivan_2025_15854453}, and GBM-TCGA \cite{GBM_TCGA}, together with one additional in-house dataset, OMG-Octo Atypical \cite{shen_2025_16780587}. Across all sources, the dataset comprised 17,664 typical mitotic figures (MFs) and 7,973 atypical mitotic figures (AMFs). The annotated MFs and AMFs in the MIDOG++ test set were reserved as an independent test cohort, while the remaining data were randomly split into training and validation subsets in an 85:15 ratio. Youden's J-Statistic was used to find the optimal threshold to maximize the balanced accuracy. 

A set of image classification architectures, including ConvNeXt and EfficientNet variants as well as UNI \cite{uni}, a vision transformer–based foundation model for pathology, were trained on the dataset. Final predictions were obtained through ensemble voting and TTA across the five best-performing models. Data augmentation consisted of random horizontal and vertical flipping (p = 0.5), RandAugment (two operations with magnitude = 9), and the colour augmentation strategy described in Track 1. 

\subsection*{Preliminary Results}
The best model for this purposes do not seem to use pre-trained foundational model but rather from-scratch trained capabilities with the ConvNext-tiny architecture, with the model size perhaps limited by the dataset size. On the test-set for MIDOG++, we find a balanced accuracy of 0.9107.

\section*{Discussion}

Our methods for the two tracks are currently under evaluation on the MIDOG 2025 preliminary test set. The inclusion of diverse datasets and the combination of the state-of-the-art architectures with our ensemble strategy is expected to yield competitive performance. Quantitative results will be presented upon the completion of the challenge evaluation period.

For Track 1, our solution differs from our latest published methodology. The SAM + CNN combination offered a comprehensive two-step model that optimizes simultaneously object detection and classification. However, it suffers from long computation time and was not adaptable to this challenge. 

For Track 2, by combining four open source datasets with an internal dataset, we created a large database that allowed us to reach a balanced accuracy of 0.9101. This score to indicate that atypical mitotic figures are significantly different from typical mitotic figures and easily distinguishable by the algorithm.

\begin{acknowledgements}
We thank the organizers of the MIDOG 2025 challenge for providing the dataset and evaluation platform.
\end{acknowledgements}


\bibliography{literature}

\end{document}